# Boundary Layer Flow over a Moving Flat Plate in Jeffrey Fluid with Newtonian Heating

Didarul Ahasan Redwan [a,*], K.M. Faisal Karim [a], M.A.H. Mamun [a] Muhammad Khairul Anuar Mohamed [b,], Syazwani Mohd Zokri [b]

[a] Department of Mechanical Engineering, Bangladesh University of Engineering and Technology, Dhaka-1000, Bangladesh.
[b] Centre for Mathematical Sciences, Universiti Malaysia Pahang, 26300 Gambang, Kuantan, Pahang, Malaysia

* Corresponding author: didar.rezwan@gmail.com

**Article history**

**Graphical abstract**

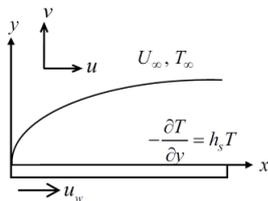

Figure 1: Physical model of the coordinate system

**Abstract**

In this paper, a numerical analysis of boundary layer flow and heat transfer in Jeffrey fluid over a moving flat plate with Newtonian Heating have been presented. The governing partial differential equations were reduced to a transformed ordinary differential equation with the help of similarity transformation. Numerical solutions were obtained for these transformed ordinary differential equation by using the Runge-Kutta-Fehlberg method. The effect on the boundary layer flow and heat transfer behaviours of various parameters such as Deborah number $\lambda_2$, relaxation time and retardation time ratio $\lambda$, Newtonian heating parameter $\gamma$, Prandtl number $Pr$ and moving plate velocity parameter $\varepsilon$ has been investigated. It is important to mention that the results obtained and reported here are impactful to the researchers working in this field and can be used in the future as a guideline and analysis context.

## INTRODUCTION

Non-Newtonian fluids have significant applications in various sectors. Such fluids have no linear relationship between deformation and stress tensor, as in a Newtonian fluids. Many fluids in real life such as custard, toothpaste, blood, petroleum and slurry are non-Newtonian fluids. Several scientists have worked with various flow models of non-Newtonian fluids (Rashaida, Bergstrom, and Sumner 2006; Sheikholeslami and Ellahi 2015; Tan and Xu 2002; Vieru, Fetecau, and Fetecau 2008; Fetecau and Fetecau 2006; Hayat and Awais 2011; Ellahi and Riaz 2010; Li et al. 2017; Hakeem, Saranya, and Ganga 2017; Sheikholeslami, Rashidi, and Ganji 2015; Attia 2008; Sahoo 2009) Jeffrey fluid model originates from Maxwell fluid model. Jeffrey model can describe both relaxation and retardation effects whereas Maxwell model can only describe relaxation effect, not retardation effect. Both relaxation and retardation effects are important in polymer industry since dilute polymer solution is a Jeffrey fluid. (Syazwani Mohd Zokri et al. 2017; 2018; S. M. Zokri et al. 2018) examined the influence of radiation and viscous dissipation on magnetohydrodynamic Jeffrey fluid and Jeffrey nanofluid over a moving plate, a stretching sheet and a horizontal circular cylinder with convective boundary conditions. They found out that the ratio of relaxation to retardation times pronounces the opposite effect to the Deborah number for both velocity and temperature profiles. (Ahmad and Ishak 2017) studied MHD Jeffrey fluid over a stretching vertical surface in a porous medium and showed that increment of magnetic parameter decreases the heat transfer rate.(Shahzad, Sagheer, and Hussain 2018) numerically simulated magnetohydrodynamic Jeffrey nanofluid flow and heat transfer over a stretching sheet considering Joule heating and viscous dissipation. They concluded that the temperature field is an increasing function of the nanoparticle volume fraction, magnetic parameter, Deborah number, Prandtl number and Eckert number. (Hayat et al. 2019) and (Aleem et al. 2020) are the most recent ones to investigate Jeffrey fluid. Hayat et al. studied melting effect in MHD stagnation point flow of Jeffrey nanomaterial and noticed intensification in flow for larger melting parameter, Deborah number and velocity ratio parameter. Maryam et al. analyzed channel flow of MHD Jeffrey fluid and found out that flow velocity increases for increasing values of thermal Grashof number, relaxation time and Jeffrey's parameter whereas it is a decreasing function of Prandtl number, porosity and Hartmann number.

In considering the flow of moving flat plate,(Sakiadis 1961) was the earliest to analyze the boundary layer flow on a constant speed moving plate.(Tsou, Sparrow, and Goldstein 1967) led an experimental study that supported Sakiadis' findings. Since then many researchers investigated boundary layer flow on a constant speed moving plate.(Erickson, Fan, and Fox 1966), (Elbashbeshy, Applied, and 2000, n.d.),(Weidman et al. 2006) and (Ishak, Yacob, and Bachok 2011) added suction or injection, temperature dependent viscosity, transpiration and radiation effect respectively to the analysis. (Anuar Mohamed et al. 2017b)used nanofluid for the analysis. They noticed that temperature profile increased because of increasing thermophoresis . (M. K.A. Mohamed et al. 2016) also used nanofluid and added viscous dissipation effect to the analysis. It was found that in the presence of viscous dissipation, the range of the plate velocity parameter reduces, which physically leads to pure conduction to occur.

The Newtonian heating boundary condition is a very sensible assumption in real world compared to classical constant wall temperature (CWT) where the wall temperature is fixed at a predefined temperature. The applications of Newtonian heating include heat exchanger, conjugate heat transfer around fins, petroleum industry, solar radiation etc. Merkin (Merkin 1994) was the earliest who considered four kinds of temperature distributions at wall and

Newtonian heating was one among them. (M. Z. Salleh, Nazar, and Pop 2009; M. Salleh et al., n.d.; M. Z. Salleh, Nazar, and Pop 2010)studied forced convection boundary layer flow at a forward stagnation point and a stretching sheet with Newtonian heating.(Muhammad Khairul Anuar Mohamed et al. 2014) investigated effects of Magnetohydrodynamic on the stagnation point flow past a stretching sheet in the presence of thermal radiation with Newtonian heating and concluded that the thermal boundary layer thickness depends strongly on magnetic parameter and thermal radiation parameter. Recent literatures on Newtonian heating are authored by (Ullah, Shafie, and Khan 2017), (Bing et al. 2017), (Al-Sharifi et al. 2017) where they investigated various effects like slip effect, radiation effect on MHD flow of different kinds of fluid on a stretching sheet.(Anuar Mohamed et al. 2017a) analyzed the effect of thermal radiation on laminar boundary layer flow over a permeable flat plate with Newtonian heating and observed that the increase of thermal radiation parameter and conjugate parameter results to the increase in wall temperature while Prandlt number does oppositely.

Motivated by the mentioned literatures, the authors intend to analyze MHD effects on boundary layer flow over a moving flat plate in Jeffrey fluid with Newtonian heating. To the best of the authors' knowledge, the study presented here is never considered before. So the obtained results are new.

**MATHEMATICAL FORMULATIONS**

Consider a horizontal moving flat plate immersed in a steady Jeffrey fluid of ambient temperature $T_\infty$ and free stream velocity $U_\infty$ as shown in Figure 1. It is assumed that $u_w(x) = \varepsilon U_\infty$ is the plate velocity where $\varepsilon$ is the plate velocity parameter. The suggested governing boundary layer equations in 2-dimensional coordinate system are (M. K.A. Mohamed et al. 2016)(S. M. Zokri et al. 2018)(S. M. Zokri et al. 2017):

$$\frac{\partial u}{\partial x} + \frac{\partial v}{\partial y} = 0, \quad (1)$$

$$u\frac{\partial u}{\partial x} + v\frac{\partial u}{\partial y} = \frac{\nu}{1+\lambda}\left[\frac{\partial^2 u}{\partial y^2} + \lambda_1\left(u\frac{\partial^3 u}{\partial x \partial y^2} + v\frac{\partial^3 u}{\partial y^3} - \frac{\partial u}{\partial x}\frac{\partial^2 u}{\partial y^2} + \frac{\partial u}{\partial y}\frac{\partial^2 u}{\partial x \partial y}\right)\right], \quad (2)$$

$$u\frac{\partial T}{\partial x} + v\frac{\partial T}{\partial y} = \alpha \frac{\partial^2 T}{\partial y^2}, \quad (3)$$

subject to the boundary conditions (M. Z. Salleh, Nazar, and Pop 2010)

$$u = \varepsilon U_\infty, \ v = 0, \ -\frac{\partial T}{\partial y} = h_s T \text{ at } y = 0,$$

$$u = U_\infty, \ T \to T_\infty \text{ as } y \to \infty, \quad (4)$$

where $u$ and $v$ are the velocity components along the $x$ and $y$ axes, respectively $\mu$ is the dynamic viscosity, $\nu$ is the kinematic viscosity, $\lambda$ is the ratio of relaxation and retardation times, $\lambda_1$ is the relaxation time, $\rho$ is a density of a based fluid, $\alpha$ is the thermal diffusivity and $T$ it local temperature while $h_s$ is the heat transfer coefficient for the Newtonian heating.

Next, in reducing the number of dependent and independent variables in Equations (1) to (3), the following similarity transformation are introduced (Bachok, Ishak, and Pop 2010)(M. Z. Salleh, Nazar, and Pop 2010)

$$\eta = \left(\frac{U_\infty}{2\nu x}\right)^{1/2} y, \quad \psi = (2U_\infty \nu x)^{1/2} f(\eta), \quad \theta(\eta) = \frac{T - T_\infty}{T_\infty}, \quad (5)$$

where $\theta$ and $\psi$ are dimensionless temperature and the stream function defined as $u = \frac{\partial \psi}{\partial y}$ and $v = -\frac{\partial \psi}{\partial x}$ which satisfy Equation (1), respectively. Therefore, $u$ and $v$ can be derived as

$$u = U_\infty f'(\eta), \quad v = -\left(\frac{U_\infty \nu}{2x}\right)^{1/2} f(\eta) + \frac{U_\infty y}{2x} f'(\eta), \quad (6)$$

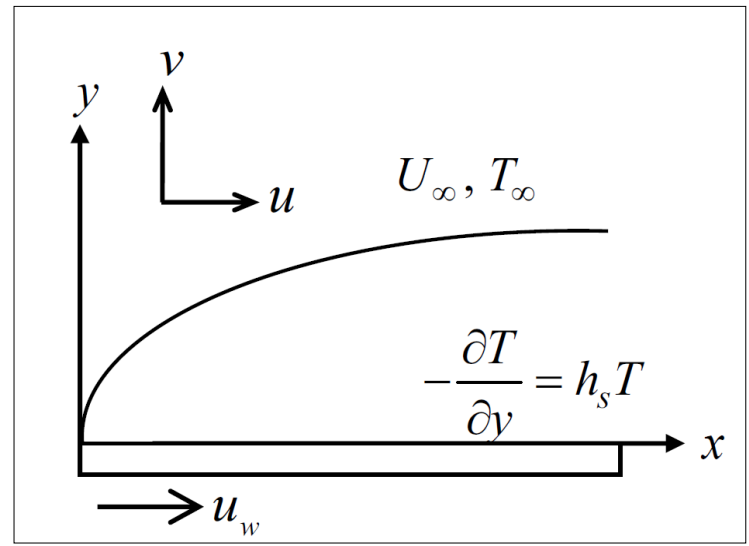

Figure 1: Physical model of the coordinate system

Substitute the Equations (5) and (6) into Equations (2) and (3), then the following transformed ordinary differential equations were obtained:

$$f''' - \frac{\lambda_2}{2} ff^{(iv)} + (1+\lambda) ff'' = 0 \quad (7)$$

$$\frac{1}{\Pr}\theta'' + f\theta' = 0, \quad (8)$$

subjected to the boundary conditions

$$f(0) = 0, \ f'(0) = \varepsilon, \ \theta'(0) = -\gamma(1+\theta(0)),$$

$$f'(\eta) \to 1, \ \theta(\eta) \to 0 \text{ as } y \to \infty. \quad (9)$$

where $\Pr = \frac{\nu}{\alpha}$, is the Prandtl number, $\lambda_2 = \frac{\lambda_1 U_\infty}{x}$ is Deborah number, and $\gamma = \left(\frac{2\nu x}{U_\infty}\right)^{1/2}$ is a conjugate parameter.

The physical quantities of interest are the skin friction coefficient $C_f$

$$C_f = \frac{\tau_w}{\rho U_\infty^2}, \quad (12)$$

where $\rho$ is the fluid density. The surface shear stress $\tau_w$, is given by(Das, Acharya, and Kundu 2015)

$$\tau_w = \frac{\mu}{1+\lambda}\left(\frac{\partial u}{\partial y} + \lambda_1 \frac{\partial^2 u}{\partial x \partial y} + v\frac{\partial^2 u}{\partial y^2}\right)_{y=0}, \quad (13)$$

with $\mu = \rho v$ and $k$ being the dynamic viscosity and the thermal conductivity, respectively. Using the similarity variables in (5) give reduced skin friction coefficient

$$C_f(2\text{Re}_x)^{1/2} = \frac{f''(0)}{1+\lambda}\left[1 - \frac{\lambda_2}{2}f'(0)\right], \quad (14)$$

where $\text{Re}_x = \frac{U_\infty x}{v}$ is the local Reynolds number.

**Table 1** Comparison between present results with previously published studies when λ=λ₂=ε=0, γ=0.1 for constant wall temperature (CWT) and $\gamma \rightarrow \infty$ for Newtonian heating (NH) case

| Pr | -θ'(0)/√2 (CWT) | | | -θ'(0) (NH) | |
|---|---|---|---|---|---|
| | (M. K.A. Mohamed et al. 2016) | (M. K.A. Mohamed et al. 2016) | Present | (M. K. A. Mohamed et al. 2017) | Present |
| 0.7 | 0.29608 | 0.292997 | 0.29278 | 0.13185 | 0.13140 |
| 0.8 | 0.30691 | 0.30724 | 0.30701 | 0.12994 | 0.12965 |
| 1 | 0.33205 | 0.332381 | 0.33215 | 0.12706 | 0.12693 |
| 5 | 0.57668 | 0.576683 | 0.57629 | 0.11398 | 0.11393 |
| 10 | 0.72814 | 0.728141 | 0.72811 | 0.11076 | 0.11070 |

**RESULTS AND DISCUSSION**

The Equations (7) and (8) along with boundary condition (9) was solved numerically by Runge-Kutta_Fehlberg method in MAPLE software. Analytical solution of these equations are highly inconvenient due to their high nonlinearity and complexity. The arbitrary values for Deborah number $\lambda_2$, Prandtl number Pr, conjugate parameter γ, plate velocity parameter ε have been set. The boundary layer thickness $\eta_\infty$ was set between 3 to 6 to get asymptotic result. The value of Pr set to 7 which is approximate Pr value of water hence provided a realistic result in analysis.

In order to validate the efficiency of the present numerical procedure, the comparison with the previously published numerical values from Mohamed et al (2017) and Zokri at el (2018) have been made and tabulated in **Table 1**. It is found that the obtained numerical values for present study are in good agreement. The numerical accuracy have been confirmed by setting the pertinent parameters as λ=λ₂=ε=0 and regards the constant wall temperature as well as Newtonian heating boundary conditions. This similarity between previously published results with present numerical values using present code gave confidence to proceed further analysis.

Next, the values of surface temperature θ(0) and reduced skin friction coefficient $C_f(2\text{Re}_x)^{1/2}$ has been tabulated in **Table 2** for different values for plate velocity parameter ε and Deborah number $\lambda_2$. From table 2, it is found that the values of θ(0) gradually declined with the increase of ε which indicates that the heat is transferred away quickly for higher values of fluid motion, thus leads to a lower plate surface temperature. The values of $C_f(2\text{Re}_x)^{1/2}$ is also decreases as ε increases for (ε<1). The values of $C_f(2\text{Re}_x)^{1/2} = 0$ as ε=1 which indicates that the fluid and the plate are in the same velocity hence, results to a no friction or a velocity differences. As ε>1, the increase of ε enhanced the values of $C_f(2\text{Re}_x)^{1/2}$ but in the opposite direction of velocity. That's explaining the negative values of $C_f(2\text{Re}_x)^{1/2}$ in table 2. As the table cross horizontally, it is noticed that the increase of $λ_2$ enhanced the values of $θ(0)$. Again, this trends occur for the case of ε<1. As ε>1, the increase of $λ_2$ reduced the values of $θ(0)$. Further, in term of $C_f(2\text{Re}_x)^{1/2}$, it is observed that a contradict trends with $θ(0)$. It is suggested that the increase of $C_f(2\text{Re}_x)^{1/2}$ is more significant with the large values of ε.

**Table 2** Values of $θ(0)$ and $C_f\left(2\text{Re}_x\right)^{1/2}$ for different values of ε and $λ_2$ when Pr= 7, λ=0.1 and γ =1.

| ε | $λ_2$=0.1 | | $λ_2$=0.3 | | $λ_2$=0.5 | |
|---|---|---|---|---|---|---|
| | $θ(0)$ | $C_f\left(2\text{Re}_x\right)^{1/2}$ | $θ(0)$ | $C_f\left(2\text{Re}_x\right)^{1/2}$ | $θ(0)$ | $C_f\left(2\text{Re}_x\right)^{1/2}$ |
| 0.1 | 11.1131 | 0.4348 | 11.7036 | 0.4230 | 12.3232 | 0.4119 |
| 0.3 | 2.6795 | 0.3847 | 2.7037 | 0.3662 | 2.7265 | 0.3486 |
| 0.5 | 1.6269 | 0.3025 | 1.6333 | 0.2813 | 1.6392 | 0.2611 |
| 1 | 0.9001 | 0.0000 | 0.9001 | 0.0000 | 0.9001 | 0.0000 |
| 2 | 0.5350 | -0.8593 | 0.5335 | -0.6444 | 0.5322 | -0.4438 |
| 3 | 0.4048 | -1.8987 | 0.4029 | -1.1698 | 0.4012 | -0.5061 |
| 5 | 0.2905 | -4.1255 | 0.2884 | -1.2753 | 0.2866 | 1.1853 |

**Figures 2** and **3** present the temperature profile θ(η) and velocity profile f'(η) for various values of ratio of relaxation time and retardation time λ, respectively. The increase of λ indicates that the relaxation time becomes more dominant and retardation time being less significant. This results to a declining of the temperature and the thermal boundary layer thickness (Abdul Gaffar, Prasad, and Reddy 2017). From **Figure 3**, it is shown that the relation between f'(η) and λ is also a negative as in Figure 2. The increase in λ causes to augmentation in shear stresses which leads to declination of fluid velocity and momentum boundary layer thickness (Syazwani Mohd Zokri et al. 2017). This situation physically increase the friction between plate surface and fluid.

**Figures 4** and 5 represent the effect of Deborah number $λ_2$ on a velocity profile f'(η) as well as a temperature profile θ(η), respectively. It is found that the increase in $λ_2$ enhanced the momentum boundary layer thickness. The values of $λ_2$ indicates the viscoelasticity property of a fluid. At a small values of $λ_2$, ($λ_2$≈0), the fluid shows more likely a Newtonian fluid characteristic and at higher values of $λ_2$, the fluid behaves like rather elastically (Hamad, AbdEl-Gaied, and Khan 2013). Deborah number $λ_2$ is linearly dependent on relaxation time that means an increase in $λ_2$ implies augmentation in relaxation time which also supports the fact the increase in velocity f'(η) as $λ_2$ increases (Awais et al. 2015).

On the other hand, from **Figure 5,** it is clearly shown that the changes in $λ_2$ gave a small influence on temperature as well as the thermal boundary layer thickness. The temperature of plate is affected marginally as observed previously in **Table.2**.

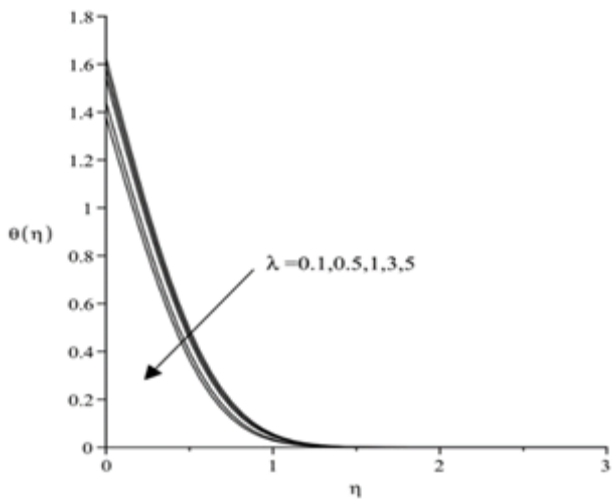

**Figure.2** Temperature profile θ(η) for various values of λ when Pr=7, $\lambda_2$=0.1, γ=1, ε=0.5

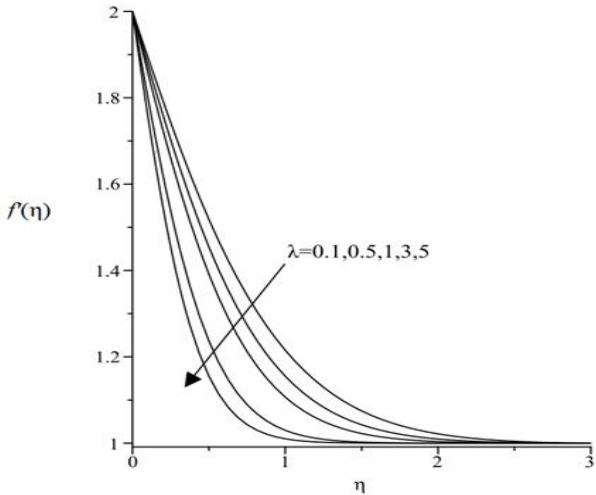

**Figure.3** Velocity profile f′(η) for various values of λ when Pr=7, γ=1, $\lambda_2$=0.1, ε=2

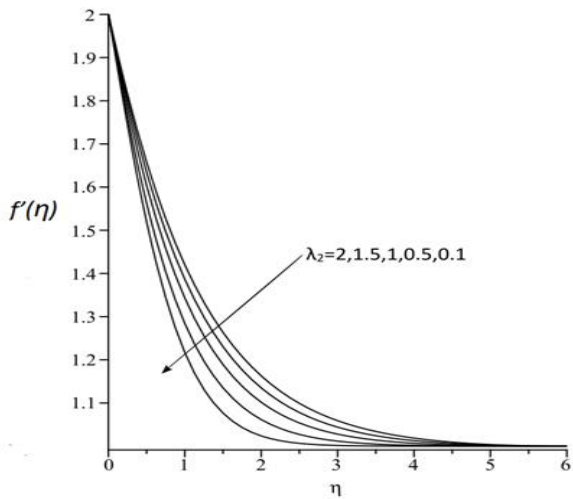

**Figure.4** Velocity profile f′(η) for various values of $\lambda_2$ when Pr=7, λ=0.1, γ=1, ε=0.5.

Next, **Figures 6 and 7** show the temperature profiles θ(η) for various values of the velocity plate parameter ε and the Prandtl number Pr, respectively. From both figures, it is found that the increase in ε and Pr results to a decreased in surface temperature θ(0) as well as the thermal boundary layer thickness. The increase of Prandtl number reduced the thermal diffusivity thus momentum diffusivity becomes more significant which leads to a declination of the thermal boundary layer thickness. Further, the increase of ε described as the fluid is moving with higher velocity thus results heat is more dissipated away from plate surface which lowering the values of θ(0).

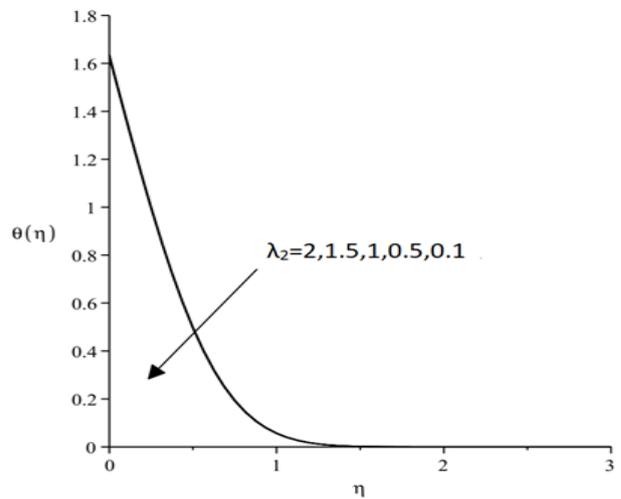

**Figure.5** Temperature profile θ(η) for various values of λ2 when Pr=7, γ=1, λ=0.1, ε=2

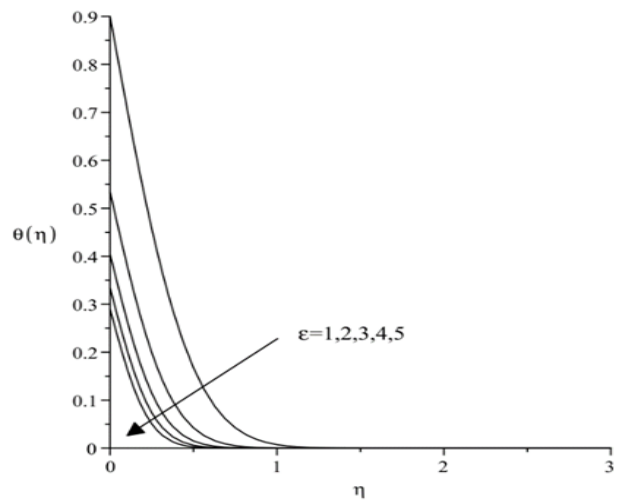

**Figure.6** Temperature profile θ(η) for various values of ε when Pr=7, λ=$\lambda_2$=0.1, γ=1

The temperature profiles θ(η) for various values of the conjugate parameter γ is illustrates in **Figure 8.** It is observed that the values of θ(0) and its thermal boundary layer thickness is increasing as γ increases. This is clear from the boundary conditions (9) where the increase in conjugate parameter proportionally results to a changes in wall temperature as well as its heat transfer coefficient. It should be mentioned that when γ tends to ∞ Newtonian heating condition becomes constant wall temperature boundary condition

**Figure.9** shows the velocity profiles $f'(\eta)$ for various values of plate velocity parameter ε. From **Figure.9** it is clearly observed that velocity of the fluid is increase with the increase of ε while the momentum boundary layer thickness effects marginally. Further, the increase of ε leads to the increase in velocity gradient which physically increase the fluid and surface friction as tabulated in **Table.2.**

Lastly, **Figures 10 and 11 present** the velocity profiles $f'(\eta)$ for various values of the Prandtl number Pr and the conjugate parameter γ, respectively. From both figures, it is observed that the fluid velocity and the momentum boundary layer thickness is not affected by the changes of γ and Pr. This is not surprisingly due to effects of velocity is clearly unrelated with both parameters as stated in Equations (8) and (9).

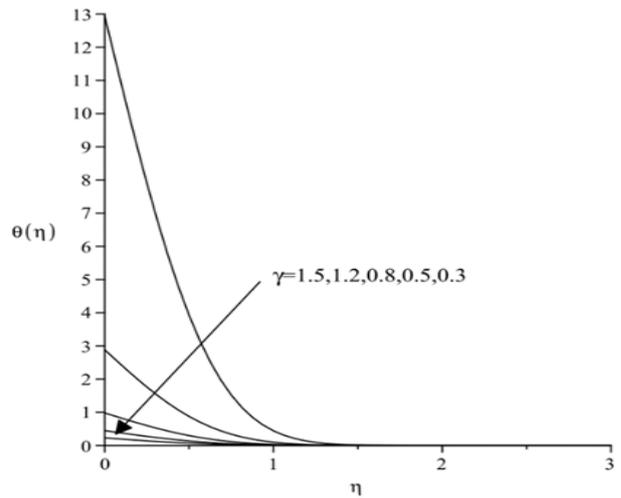

**Figure.8** Temperature profile θ(η) for various values of γ when Pr=7,λ=λ$_2$=0.1,ε=2

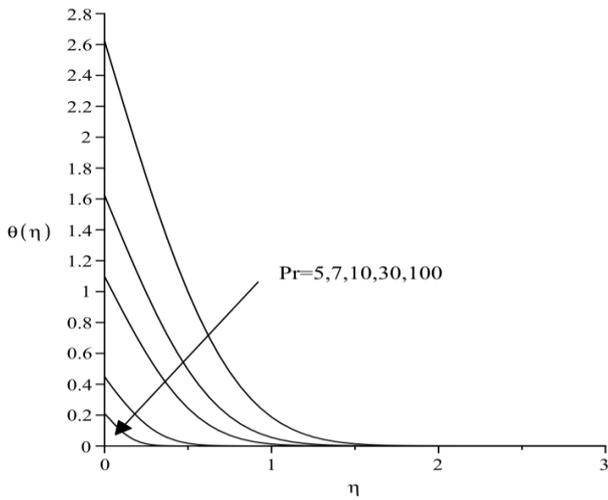

**Figure.7** Temperature profile θ(η) for various values of Pr when γ=1,λ=λ$_2$=0.1,ε=0.5.

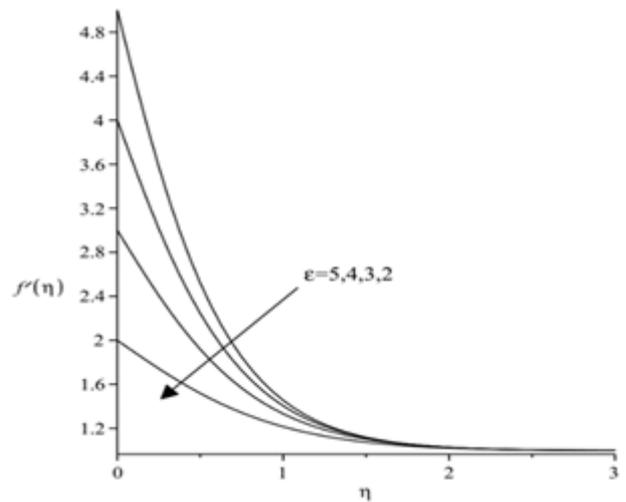

**Figure.9** Velocity profile f'(η) for various values of ε when Pr=7,λ=λ2=0.1,γ=1

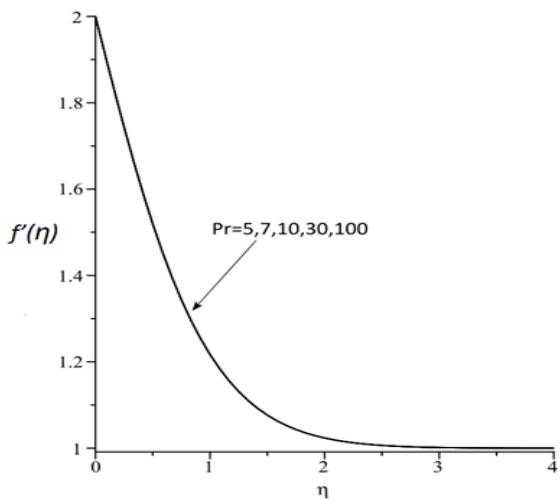

**Figure.10** Velocity profile $f'(\eta)$ for various values of Pr when $\gamma=1, \lambda=\lambda_2=0.1, \varepsilon=2$

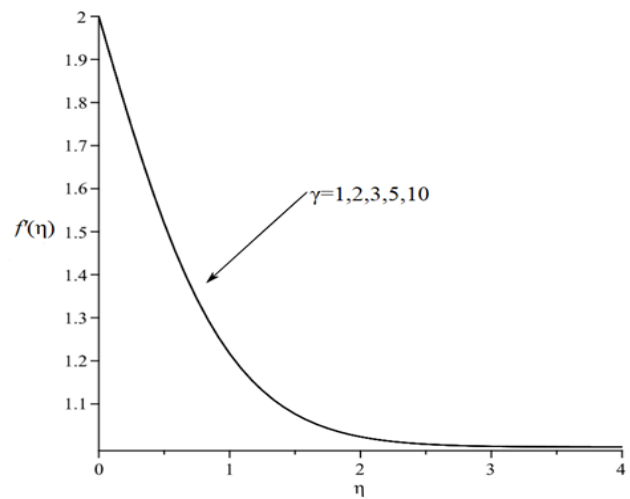

**Figure.11** Velocity profile $f'(\eta)$ for various values of $\gamma$ when Pr=7, $\lambda=\lambda_2=0.1, \varepsilon=0.5$.

## CONCLUSION

- The increase in Deborah number has increased marginally the surface temperature at $\varepsilon<1$, while decrease as $\varepsilon>1$. The skin friction coefficient act contrary.
- The skin friction coefficient tends to 0 as $\varepsilon$ approaches 1.
- Increase in $\lambda$ leads to declination both in thermal and momentum boundary layer thicknesses.
- Temperature profile declines with augmentation of Pr and $\varepsilon$ but increases with an increase in $\gamma$.
- Pr and $\gamma$ do not have any significant impact on velocity profile.

## ACKNOWLEDGEMENT

The authors would like to express deep gratitude to Department of Mechanical Engineering, Bangladesh University of Engineering and Technology (BUET) for providing continuous encouragement throughout this investigation.